\documentclass[aps,twocolumn,showpacs]{revtex4}
\usepackage{graphicx}
\usepackage{times}
\begin{document}
\preprint{full\_RT\_NBEPL\_ZnOGa\_JAP5.tex, JAP}
\title{Spectral shape analysis of ultraviolet luminescence in \textit{n}-type ZnO:Ga}
\author{T. Makino} 
\altaffiliation[Present address: ]{Department of Material Science, University of Hyogo, Kamigori-cho, Hyogo 678-1297, Japan}
\email[electronic mail: ]{makino@sci.u-hyogo.ac.jp}
\author{Y. Segawa}
\affiliation{Photodynamics Research Center, RIKEN (Institute of Physical and Chemical Research), Sendai 980-0845, Japan}
\author{S. Yoshida}
\author{A. Tsukazaki}
\author{A. Ohtomo}
\author{M. Kawasaki}
\altaffiliation[Also at: ]{Combinatorial Materials Exploration and Technology, National Institute for Materials Science, Tsukuba 305-0044, Japan}
\affiliation{Institute for Materials Research, Tohoku University,
Sendai 980-8577, Japan}
\author{H. Koinuma}
\affiliation{National Institute for Materials Science, 1-2-1 Sengen, Tsukuba, Ibaraki, 305-0047, Japan}
\date{\today}
\begin{abstract}
Thin films of laser molecular-beam epitaxy grown \textit{n}-type Ga-doped ZnO were investigated with respect to their optical properties.
Intense room-temperature photoluminescence (PL) in the near-band edge (NBE) region was observed. Moreover, its broadening of PL band was significantly larger than predicted by theoretical results modeled in terms of potential fluctuations caused by the random distribution of donor impurities. In addition, the lineshape was rather asymmetrical. To explain these features of the NBE bands, a vibronic model was developed accounting for contributions from a series of phonon replicas.
\end{abstract}
\maketitle
Recently, optical properties of near-band-edge (NBE) recombination in ZnO epitaxial layers have been extensively investigated. Compared to undoped ZnO, the properties of heavily-donor doped ZnO, such as phonon interactions, are not well understood. In our previous work, we reported the observation of intense NBE photoluminscence (PL) for ZnO:Ga at room temperature (RT). Figure~1(a) shows their PL spectra at four different doping levels (dashed lines). The experimental procedures~\cite{makino_int_Ga} were identical to those adopted in our previous study and their details were reported elsewhere~\cite{makino19}. The shape of the PL band is asymmetrical. The linewidths of the PL increase from 154 to 195~meV with an increase in Ga concentration ($n_{Ga}$). Previously, we estimated the ``theoretical'' broadening for the PL in terms of potential fluctuations caused by the random distribution of donor impurities (cf. Fig.~3 of Ref.~\onlinecite{makino_int_Ga}). These values are also listed in Table~I. However, the calculated widths ($\sigma_1$) were smaller than the experimentally observed broadening.
\begin{figure}
\caption{(Color online) (a) Room-temperature photoluminescence spectra
(dashed curves) of \textit{n}-type ZnO doped with different Ga concentrations.
Also shown are the results of the fit to the data using a `vibronic' model (solid lines). (b) The lowermost PL curve of frame (a) with individual contribution
of the emission lines (dash-dotted lines).}
\label{raw-data}
\includegraphics[width=.52\textwidth]{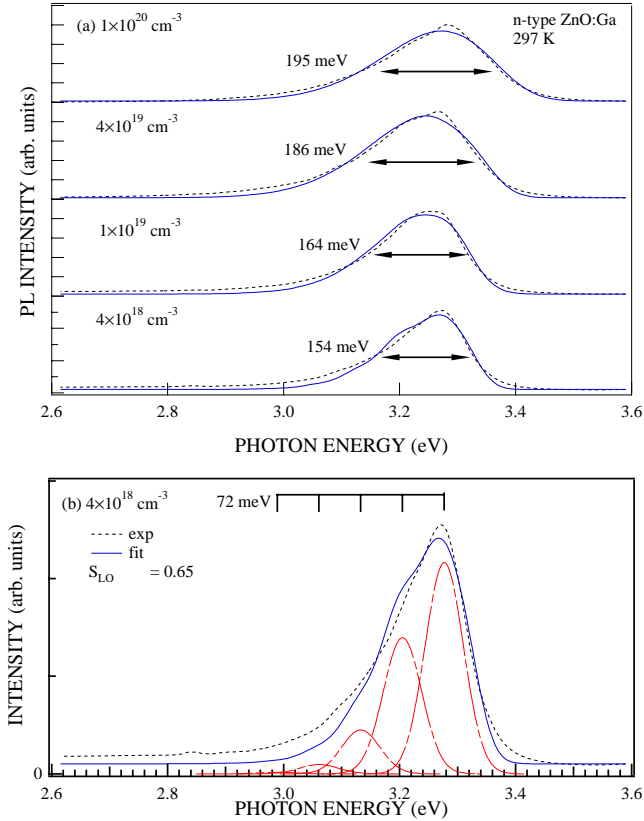}
\end{figure}
Table I: Sample specification and main characteristics for the four \textit{n}-ZnO samples.
Ga concentration ($n_{Ga}$), theoretical broadening $\sigma_1$, broadening
evaluated from the Monte Carlo simulation $\sigma_2$, zero-phonon peak energy $E$, and
the deduced Huang-Rhys factor $S$. \\[1cm]
	\begin{ruledtabular}
		\begin{tabular}{ccccc}
$n_{\rm Ga}$ & $\sigma_1$ & $\sigma_2$ & $E$ & $S$\\
(cm$^{-3}$)&(meV)&(meV)& (eV) & \\
\hline

8.0$\times10^{18}$ & 77.7 & 73.5 & 3.277 & 0.64\\
2.0$\times10^{19}$ & 97.1 & 115 &3.272& 0.78\\
8.0$\times10^{19}$ & 150 & 139 &3.297& 1.1\\
1.5$\times10^{20}$ & 188 & 170 &3.345& 1.4\\
\hline
		\end{tabular}
	\end{ruledtabular}

These theoretical widths might be exceedingly underestimated because the model might be too
much simplified. We here adopt another approach properly taking the microscopic fluctuation of donor concentration into account; the thermalization redistribution model developped by Zimmermann \textit{et~al}~\cite{zimmermann1}. According to this model, there is a relationship between the Stokes shift and the full-width at half maximum (FWHM) of PL. One can estimate the energy scale of the band potential profile fluctuation from the Stokes shift and predict consistently the FWHM. Figure~2 shows a Stokes shift of the luminescence plotted against the gallium concentration. Obviously, the shift energy increases with incorporation of the donor dopants.
\begin{figure}
	\caption{The PL Stokes shifts (closed squares) plotted against the Ga concentration.}
\includegraphics[width=.52\textwidth]{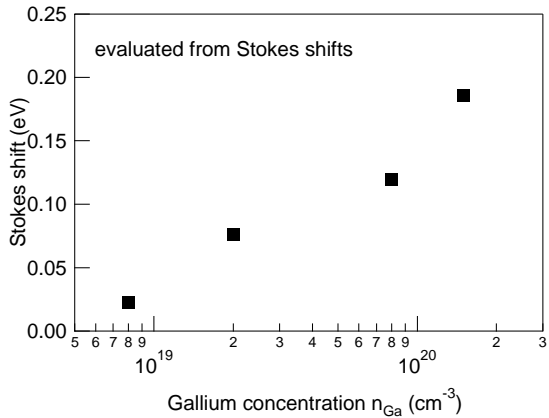}
	\label{Stokes}
\end{figure}

At sufficiently high temperatures, the above-mentioned relationship is indicated as $E_{Stokes}=-\sigma^2/kT$. For example, the Stokes shift for the sample with the $n_{Ga}$ of 8.0 $\times 10^{18}$~cm$^{-3}$ is 22~meV, leading to $\sigma $ of 24~meV. It also yields the temperature independent value of 2$\sigma \sqrt(\ln4) \simeq 46$~meV in the FWHM. This can be also supported by the Monte Carlo simulation. We now simulate a PL spectrum.

Miller-Abraham's rate for phonon-assisted exciton tunneling between the initial and final states $i$ and $j$ with the energies of $E_i$ and $E_j$ was adopted to simulate the hopping events of excitons or of photocreated carriers:
\begin{equation}
	\nu_{ij}=\nu_0 \exp \left(- \frac{2r_{ij}}{\alpha} - \frac{(\varepsilon_i-\varepsilon_j+|\varepsilon_i-\varepsilon_j|)}{2kT} \right).
\end{equation}
Here $r_{ij}$ is the distance between the localized states, $\alpha$ is the decay length of the wave function, and $\nu_0$ is the attempt-escape frequency. Hopping was simulated over a randomly generated set of localized states with the sheet density of $N$. Dispersion of the localization energies was assumed to be in accordance with a Gaussian distribution,
\begin{equation}
	g(\varepsilon ) \propto \exp(- (\varepsilon-E_0)^2/2\sigma^2).
\end{equation}
With the peak positioned at the mean excitonic energy $E_0$ and the dispersion
parameter (the energy scale of the band potential profile fluctuation) $\sigma$. For each generated exciton, the hopping process terminates by recombination with the probability $\tau_0^{-1}$ and the energy of the localized state, where the recombination has taken place is scored to the emission spectrum.

In addition, the thermal broadening effect was implemented by convoluting the above spectrum with a Gaussian curve, which describes the thermal part with a width of $1.8 k_B T$~\cite{schubert2}. The nomenclatures ($k_B$ and $T$) take their conventional meanings. Because the above equation contains the term of thermal activation, at a glance, it seems that it is unnecessary to perform the convolution. However, because Kazlauskas \textit{et~al.} adopted this treatment, we follow it in this work~\cite{kazl1}.
Meanwhile, the overall FWHM of the PL band is given:
\begin{equation}
\textrm{FWHM (}\sigma_2\textrm{)} \approx [(2\sigma \sqrt(\ln4))^2+(1.8 k_B T)^2]^{1/2}.
\end{equation}
\begin{figure}
	\caption{Simulated PL spectrum under stationary excitation. The Ga concentration is shown in the figure.}
	\label{simulated}
\includegraphics[width=.52\textwidth]{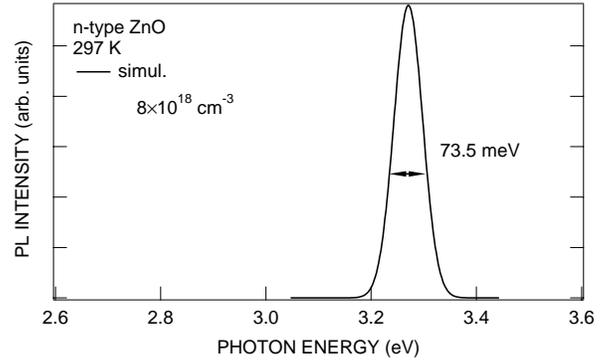}
\end{figure}
The solid line in Fig.~3 represents the simulated result obtained for the following values of parameters: $N \alpha^3 =0.1$, $\tau_0 \nu_0=10^4$, and $\sigma \simeq 24$~meV. The line-shape of resulting spectrum was independent of the choices of the $N \alpha^3$ and $\tau_0 \nu_0$ parameters. The simulated curve could be well approximated by the Gaussian function, which is not surprising if considering the measurement temperature (300~K). We summarize the FWHMs ($\sigma_2$) of the simulated PL spectra for the samples at four different doping levels in Table~I.

The simulated broadening ($\sigma_2$) is still smaller than that of experiments. This discrepancy is thought to be due to the contribution of the phonon replicas. In addition, the simulated line-shape is rather symmetrical, which is in great contrast to the asymmetrical shape observed in the experimental spectra. In compensated semiconductors, even a low-temperature emission spectrum shows very broad feature. Its line-shape is sometimes asymmetrical. Kuskovsky and his co-workers~\cite{kuskovsky1} have submitted a theory reproducing such an asymmetrical line-shape. On the other hand, our experiments were performed at room temperature. Therefore, it is natural to think that the shape of a unresolved zero-phonon band is rather symmetrical because of the thermalization effects. We did not estimate the width and the asymmetry of our PL using their theory because, unfortunately, it is applicable only to the case of extremely low temperatures and because our samples are not very severely compensated~\cite{makino_int_Ga}. We think that the zero-phonon peak and its phonon replicas are not spectrally resolved well in the spectrum because intensity of a one-phonon replica is comparable to that of a zero-phonon band, leading to the larger experimental broadening.

The typical consequence of the interaction with phonons is the appearance of phonon-assisted emissions in a PL spectrum. We infer that the coupling with longitudinal-optical (LO) phonon is very efficient in this highly polar material. The reported coupling constant of ZnO is nearly four times stronger than that of GaN and about eight times stronger than that of ZnSe~\cite{makino8}.

For simplicity, we took only the terms of longitudinal optical (LO) phonons into account~\cite{dukeandmahan,bkmeyer-rev,reynolds-green}, whose interaction is the strongest. In this case, the transition energy of the replicas will be $\hbar \omega = E_{\rm ZPL} -  \eta_{\rm LO} E_{\rm LO}$, where $E_{\rm LO}$=72~meV in ZnO, $E_{\rm ZPL}$ is the energy of the zero-phonon peak and $\eta_{\rm LO} $ is an integer.

Each individual emission line in the spectrum was modeled by a Gaussian function with linewidth $\sigma_1$ (cf. Table~I), which can be justified by the Gaussian-like lineshape observed in our simulated spectrum. The values of $\sigma_1$ were determined from the above-mentioned theory because these values are larger than those of the Monte Carlo simulation, $\sigma_2$. The probability of a given phonon emission is proportional to $(S_{\rm LO}^{\eta {\rm LO}}/ \eta_{\rm LO}!)$, where $S_{\rm LO}$ is the Huang-Rhys factor for the LO phonons and $\eta_{\rm LO}$ the number of phonons emitted in a transition. The solid lines in Figs.~\ref{raw-data}(a) and (b) were obtained by summing all of the emission lines using an appropriate set of parameters. The donor-impurity concentrations ($n_{\rm Ga}$), zero-phonon peak energies ($E$), and \textit{S}-factors are compiled in Table~I. Five dash-dotted curves in Fig.~\ref{raw-data}(b) denote the respective contributions whose peak positions are equidistant by the $E_{\rm LO}$ of 72~meV. The important conclusion from this fit is that the asymmetrical and broad PL band can be explained by the contribution of LO phonon replicas. It should be noted that the Huang-Rhys factor ($S_{\rm LO}$) depends on the ionic nature of semiconductors and on the separation of Fourier transformed charge distributions. This sometimes deviates from the bulk value~\cite{dukeandmahan,bkmeyer-rev,reynolds-green}. Very recently, Shan \textit{et~al.}~\cite{w_shan1} reported that the RT PL spectrum in undoped ZnO is dominated by the phonon replicas of the free exciton transition with the maximum at the first LO phonon replica. This corresponds to the \textit{S}-factor exceeding unity. In addition, in the undoped ZnO, the zero-phonon emission becomes weaker than 1LO-phonon-assisted annihilation processes at elevated sample temperatures.

Even if the values of $\sigma_1$ are comparable with the optical phonon energy $E_{LO}$, due to the high measurement temperature, we believe it is probable that the radiative transition would go via phonon-assisted path while the impurity levels are available at the same energies.

We see the shoulders at the 1LO phonon replica position in both the fitting and experimental results, particularly in the first two cases. The ``experimental'' shoulder is less prominent than that of fit. The reason of this is not clear
at this moment but we speculate that this is because of our assumption that
the width of the phonon replicas is same with that of the zero-phonon band.
We think that the broader replicas are more realistic.

In summary, gallium was successfully applied as a source of donor impurities in the laser molecular-beam epitaxy growth of \textit{n}-type ZnO. We presented spectroscopic measurements to study the evolution of spectra in \textit{n}-type ZnO along with Ga concentrations. Asymmetric and broad photoluminescence spectra were fitted to a model that takes into account phonon replicas, in which the linewidths of each individual emissions have been determined mainly from the concentration fluctuation of dopants.

\textbf{Acknowledgments}---One of the authors (T.~M.) is thankful to K.~Saito of the University of Tokyo, Japan for providing his simulation algorithm. This work was partially supported by MEXT Grant of Creative Scientific Research 14GS0204, the Asahi Glass Foundation, and the inter-university cooperative program of the IMR, Japan.

\end{document}